\documentclass[11pt]{article}
\usepackage{graphicx,amsmath,bm, amsthm,mathrsfs,amssymb, braket, verbatim}
\usepackage[usenames]{color}
\usepackage{ulem,mathtools}
\usepackage{authblk}
\usepackage{cite}
\usepackage{tcolorbox}

\newcommand{\ee}{\end{equation}}

\newcommand{\reff}[1]{(\ref{#1})}
\newcommand{\beq}{\begin{equation}}
\newcommand{\eeq}[1]{\label{#1}\end{equation}}
\newcommand{\beqa}{\begin{eqnarray}}
\newcommand{\eea}{\end{eqnarray}}
\newcommand{\eeqa}[1]{\label{#1}\end{eqnarray}}
\newcommand{\beg}{\begin{equation*}}
\newcommand{\eeg}{\end{equation*}}

\newcommand{\bsplit}{\begin{split}}
\newcommand{\esplit}{\end{split}}

\usepackage{subfig} 
\usepackage{circuitikz} 
\usepackage[capposition=bottom]{floatrow} 
\usepackage{epigraph} 
\usepackage{appendix}
\usepackage{rotating}
\allowdisplaybreaks

\title{Relativistic corrections to Landau levels in the presence of a parallel linear electric field}

\author[]{Yann Audin\thanks{yaudin13@ubishops.ca}}
\author[]{Ariel Edery\thanks{aedery@ubishops.ca}}

\affil[]{Department of Physics and Astronomy, Bishop's University, 2600 College Street, Sherbrooke, Qu\'{e}bec, Canada, J1M 1Z7.\vspace{1em}}

\begin{document}
\date{}
\maketitle
\begin{abstract}
We consider an electron moving under a constant magnetic field (in the z-direction) and a \textit{linear} electric field parallel to the magnetic field above the z=0 plane and anti-parallel below the plane. Two frequencies characterize the system: the cyclotron frequency $\omega_c$ corresponding to motion along the x-y plane and associated with the usual Landau levels, and a second frequency $\omega_z$ corresponding to motion along the z-direction. In previous work, the non-relativistic energies of this system were obtained, and it was shown that an extra degeneracy (beyond the Landau degeneracy) occurs when the ratio $\text{w}=\omega_c/\omega_z$ is rational. In this paper, we use Dirac's equation to obtain compact formulas for the first and second order relativistic corrections to this system via perturbation theory. The formulas are expressed in terms of the two frequencies $\omega_c$ and $\omega_z$, and two quantum numbers, $n$ and $n_z$, both of which are non-negative integers. The first order correction is negative and lowers the original energies. We plot the energy (zeroth plus first order) versus the ratio $\text{w}$ and there are degeneracies at all points where lines intersect. However, the degeneracy does not occur at the same $\text{w}$ as before. To illustrate this, we show how the first order correction splits the energy levels for the case $\omega_c=\omega_z$.          
\end{abstract}
\setcounter{page}{1}
\newpage
\section{Introduction}\label{Intro}
An electron moving in two dimensions (x-y plane) under a constant magnetic field (in the z-direction) has discrete energies known as Landau levels. The energies, neglecting spin, are simply those of the one-dimensional harmonic oscillator with cyclotron frequency $\omega_c$. In a previous paper \cite {EA} we added a \textit{linear} electric field to the Landau scenario of a constant magnetic field. The electric field was parallel to the magnetic field above the z=0 plane and anti-parallel to the magnetic field below the z=0 plane. We obtained the non-relativistic energies and degeneracy of this system. In this paper, we calculate the first and second order relativistic corrections to this scenario using Dirac's equation and perturbation theory. 

In the absence of an electric field, it is possible to obtain an exact expression for the relativistic Landau levels using Dirac's equation (for a treatment of this problem see \cite{Maggiore}). Our goal here is not to find corrections to that result due to the perturbations of an electric field. In other words, our zeroth order result is not the relativistic Landau levels with magnetic field only, but the non-relativistic result found in \cite{EA} that includes both the magnetic and electric fields. 

The literature contains studies of relativistic Landau levels in a variety of contexts. Relativistic landau levels for a spinless particle have been studied in a rotating cosmic string spacetime using the Klein-Gordon equation \cite{Muniz}. The finite-temperature relativistic Landau problem and the relativistic quantum Hall effect were investigated in \cite{Bene}, where for the latter, crossed electric and magnetic fields were used.  Relativistic Landau quantization of a neutral particle with a permanent magnetic dipole moment was studied in \cite{Bakke} where the applied electric field was non-uniform and confined to the x-y plane (with magnetic field in the $z$-direction). Landau level spectroscopy of relativistic fermions in the $\text{Bi}_2\text{Te}_3$ three-dimensional topological insulator was studied in \cite{Wolos} where intraband transitions of Landau levels of relativistic fermions were observed. Dirac cat states in relativistic Landau levels were studied in \cite{Bermudez}. They showed that a relativistic version of Schrodinger cat states, Dirac cat states, can be built in relativistic Landau levels when an external magnetic field couples to a relativistic spin $1/2$ charged particle.  

In the above literature, whenever an electric field is present, it is applied in the x-y plane. In this work, a linear electric field is applied in the $z$-direction. At the zeroth order (non-relativistic level), this leads the particle to oscillate harmonically about $z=0$ along the z-direction with frequency $\omega_z$. We therefore obtain a three-dimensional system with two characteristic frequencies: $\omega_c$ and $\omega_z$. We obtain compact formulas for the first and second order relativistic corrections to this system. They are expressed in terms of two non-negative quantum numbers $n$ and $n_z$, and the two frequencies $\omega_c$ and $\omega_z$. The first order correction is negative and lowers the energies. We plot the energy (zeroth plus first order) versus the ratio $\text{w}=\omega_c/\omega_z$. We see that an extra degeneracy occurs at all points where lines intersect. However, the degeneracies do not occur exactly at the same value of $\text{w}$ as in the non-relativistic case. To demonstrate this, we draw an energy level diagram that shows how the first order correction splits the energy levels for the case when $\omega_c=\omega_z$.     

\section{Dirac's equation for a charged particle moving in a linear electric field parallel to a uniform magnetic field}

The Dirac equation for an electron of charge $-e$ and mass $m_e$ moving in an external electromagnetic field $A_{\mu}$ is given by
\beq
[\gamma^{\mu}(i \hbar\,\partial_{\mu} +e A_{\mu}) -m_e]\Psi=0\,.
\eeq{Dirac}
Let $\psi_L$ and $\psi_R$ be left-handed and right-handed Weyl spinors respectively. We work in the standard representation where $\Psi= \dfrac{1}{\sqrt{2}}\begin{pmatrix} \psi_R +\psi_L\\\psi_R -\psi_L\end{pmatrix}=\begin{pmatrix} \Phi \\\chi\end{pmatrix}$ 
and the gamma matrices are given by 
\beq
\gamma^0=\left( {\begin{array}{cc} 1 & 0 \\0 & -1 \\ \end{array} } \right)\quad,\quad \gamma^i=
  \left( {\begin{array}{cc} 0 & \sigma^i \\-\sigma^i& 0 \\ \end{array} } \right)\,.
\eeq{gamma}
The $\sigma^i$ (i=1,2,3) are the Pauli matrices. The Dirac equation \reff{Dirac} can then be expressed as the following two equations:
\begin{align}
(i \hbar\,\partial_0 +e A_0-m_e) \Phi + \sigma^i(i \hbar\,\partial_i +e A_i) \chi &=0 \label{phi}\\
(i \hbar\, \partial_0 +e A_0 +m_e) \chi + \sigma^i(i \hbar\, \partial_i +e A_i) \Phi&=0\,.\label{chi}
\end{align}
We look for stationary solutions $\Phi({\bf x},t)=\Phi({\bf x})e^{-i E t/\hbar}$ and $\chi({\bf x},t)=\chi({\bf x})e^{-i E t/\hbar}$. Substituting this into the above two equations we obtain
\begin{align}
(E +e A_0-m_e) \Phi &= \bm{\sigma}.({\bf p} +e {\bf A}) \chi  \label{phi3}\\
(E +e A_0 +m_e) \chi &= \bm{\sigma}.({\bf p} +e {\bf A}) \Phi\label{chi3}
\end{align}
where ${\bf p} =-i \hbar {\bf \nabla}$ was used. Substituting $\chi$ from \reff{chi3} into \reff{phi3} yields the equation 
\begin{align}
((E +e A_0)^2-m_e^2) \Phi&= [\bm{\sigma}.({\bf p} +e {\bf A})]^2 \Phi \label{G1}\\
&=[({\bf p} +e {\bf A})^2 +e \bm{\sigma}.{\bf B}]\Phi \label{G2}
\end{align}
where in going from \reff{G1} to \reff{G2} we used the algebraic identity $\sigma^i\sigma^j=\delta^{ij} +i\epsilon^{ijk}\sigma^k$ and $\epsilon^{ijk} \nabla^i A^j=B^k$ where $B^k$ is the $k$ component of the magnetic field ${\bf B}$.  We can express the above formally in terms of a square root operator
\begin{align}
E  \Phi &=  (\sqrt{m_e^2+ ({\bf p} +e {\bf A})^2 +e \bm{\sigma}.{\bf B}} -e A_0) \Phi\\
&=(m_e\sqrt{1+ \dfrac{({\bf p} +e {\bf A})^2}{m_e^2} + \dfrac{e \bm{\sigma}.{\bf B}}{m_e^2}} -eA_0) \Phi
\end{align} 
We now assume the kinetic energy and spin term are much less than the rest mass and we perform a binomial expansion,     
\begin{align}
E  \Phi = \Big[ m_e\Big(1+ \dfrac{1}{2} \Big(\dfrac{({\bf p} +e {\bf A})^2}{m_e^2} &+ \dfrac{e \bm{\sigma}.{\bf B}}{m_e^2}\Big)-\dfrac{1}{8} \Big(\dfrac{({\bf p} +e {\bf A})^2}{m_e^2} + \dfrac{e \bm{\sigma}.{\bf B}}{m_e^2}\Big)^2 \nonumber\\&+ \dfrac{1}{16} \Big(\dfrac{({\bf p} +e {\bf A})^2}{m_e^2} + \dfrac{e \bm{\sigma}.{\bf B}}{m_e^2}\Big)^3 +...\Big) -e A_0 \Big]\Phi\\
= (H^{(0)} +H^{(1)} + H^{(2)} +...) \Phi
\end{align}
where 
\begin{align}
&H^{(0)}= m_e+ \dfrac{1}{2} \Big(\dfrac{({\bf p} +e {\bf A})^2}{m_e} + \dfrac{e \bm{\sigma}.{\bf B}}{m_e}\Big) -e A_0\label{H0} \\
&H^{(1)}=-\dfrac{m_e}{8} \Big(\dfrac{({\bf p} +e {\bf A})^2}{m_e^2} + \dfrac{e \bm{\sigma}.{\bf B}}{m_e^2}\Big)^2 \label{H1}\\
&H^{(2)}= \dfrac{m_e}{16} \Big(\dfrac{({\bf p} +e {\bf A})^2}{m_e^2} + \dfrac{e \bm {\sigma}.{\bf B}}{m_e^2}\Big)^3 \label{H2}
\end{align}
We now consider a uniform magnetic field ${\bf B}=B_0 \hat{z}$ with vector potential given in the Landau gauge as ${\bf A} = B_0 x \hat{y}$, and a linear electric field ${\bf E}= k z \hat{z}$ derived from the scalar potential $\phi=A_0=-k z^2/2$ where $k$ is a positive constant. We therefore obtain that 
\beq
H^{(0)}= m_e + \dfrac{p_x^2}{2 m_e} +\dfrac{(p_y+e B_0 x)^2}{2m_e} + \dfrac{p_z^2}{2m_e} + \dfrac{e}{2m_e}\sigma_z B_0 +\dfrac{e}{2}kz^2 \,.
\eeq{H_0}
Except for the rest mass $m_e$, $H^{(0)}$ is the non-relativistic Hamiltonian for an electron moving in a uniform magnetic field and a parallel linear electric field \cite{EA}. Therefore $H^{(1)}$ and $H^{(2)}$ are relativistic perturbations to the Hamiltonian $H^{(0)}$. The Hamiltonian $H^{(0)}$ commutes with $p_y$, $\sigma_z$ and also $H_z=\dfrac{p_z^2}{2m_e} + \dfrac{e}{2}\,k\,z^2$ ; therefore these operators share a common eigenfunction $\psi_0\,$, 
\beq
H^{(0)} \,\psi_0 = E^0 \,\psi_0\quad;\quad H_z \psi_0=\varepsilon_z \psi_0 \quad;\quad \sigma_z \,\psi_0=\pm \hbar\,\psi_0 \quad;\quad
 p_y \,\psi_0=\hbar \,k_y\,\psi_0 \,.
\eeq{H3}
It is now convenient to rewrite $H^{(0)}$ in the form 
\beq
H^{(0)}=  m_e + \dfrac{p_x^2}{2 m_e} +\dfrac{1}{2}m_e\omega_c^2(x-x_0)^2 \pm \dfrac{1}{2} \hbar\omega_c + \dfrac{p_z^2}{2m_e} + \dfrac{1}{2}m_e\omega_z^2 \,z^2
\eeq{H_0}
where
\beq
\omega_c=\dfrac{e\,B_0}{m_e}
\eeq{wc}
is the usual cyclotron frequency,
\beq
\omega_z= \sqrt{\dfrac{ek}{m_e}}
\eeq{wz} 
is the frequency of oscillations in the z-direction and 
\beq
x_0=-\dfrac{\hbar \,k_y}{e\,B_0}\,.
\eeq{x0}
The quantity $\dfrac{p_x^2}{2 m_e} +\dfrac{1}{2}m_e\omega_c^2(x-x_0)^2$ is the operator of a quantum harmonic oscillator centered at $x=x_0$ with eigenvalue $(n_c + 1/2) \hbar \omega_c$ where $n_c=0,1,2,3,...$. We now include the spin contribution and define the operator $H_x$ as   
\beq
H_x=\dfrac{p_x^2}{2 m_e} +\dfrac{1}{2}m_e\omega_c^2(x-x_0)^2 \pm \dfrac{1}{2} \hbar\omega_c\,.
\eeq{Hx}
Then $H_x\,\psi_0= [(n_c + \dfrac{1}{2}) \hbar \omega_c\pm \dfrac{1}{2} \hbar\omega_c]\,\psi_0= n \,\hbar \,\omega_c \,\psi_0$ where $n=0,1,2,3...$. There is now a twofold spin degeneracy for every $n$ except $n=0$ which has a spin degeneracy of unity. The quantity $H_z$ is also the operator for a quantum harmonic oscillator such that  $H_z \,\psi_0= \Big(n_z + \dfrac{1}{2}\Big) \hbar \,\omega_z\, \psi_0$ with $n_z=0,1,2,3,...$. The non-relativistic energy $E^0$ corresponding to the Hamiltonian $H^{(0)}$ is thus given by 
\beq
E^0= m_e+ n \,\hbar \,\omega_c + \Big(n_z + \dfrac{1}{2}\Big) \hbar \,\omega_z \,.
\eeq{E0}
Except for the rest mass $m_e$, this is the non-relativistic energy obtained in \cite{EA}. In the sections that follow we will refer to $\psi_0$ as the eigenstate $\ket{n,n_z}$.
\section{First order relativistic correction}
The first order relativistic correction to $E^0$ is given by 
\beq
E^{(1)}=\bra{n,n_z} H^{(1)} \ket{n, n_z}
\eeq{E1}
where $H^{(1)}$ is given by \reff{H1}. The operator $H^{(1)}$ can be expressed in terms of $H_x$ and $p_z$,
\begin{align}
H^{(1)} &= -\dfrac{1}{2m_e}\Big(H_x + \dfrac{p_z^2}{2m_e}\Big)^2\nonumber\\
&=-\dfrac{1}{2m_e}\Big(H_x^2 + H_x\,\dfrac{p_z^2}{m_e}+ \dfrac{p_z^4}{4 m_e^2}\Big)\,.
\label{H1A}
\end{align}
Note that $H_x$ and $ p_z^2$ commute. We know that 
\beq
H_x \ket{n,n_z} = n \,\hbar \,\omega_c \ket{n,n_z}\,.
\eeq{HxA}
To evaluate how the powers of $p_z$ act on the eigenstate $\ket{n,n_z}$ we define the following creation and annihilation operators 
\begin{align}
a^{\dagger}&=\sqrt{\frac{m_e \omega_z}{2\hbar}}(z-\frac{i}{m_e \omega_z} p_z) \nonumber\\
a&=\sqrt{\frac{m_e \omega_z}{2\hbar}}(z+\frac{i}{m_e \omega_z} p_z)\,.
\end{align}
They obey the commutation relation $[a,a^{\dagger}]=1$. We have that $a^{\dagger}\ket{n,n_z}=\sqrt{n_z+1}\ket{n, n_z+1}$ and $a\ket{n,n_z}=\sqrt{n_z}\ket{n,n_z-1}$.
The operator $p_z$ can be expressed as 
\beq
p_z= i \sqrt{\frac{\hbar \,m_e \omega_z}{2}} (a^{\dagger}-a)\,.
\eeq{pz}
We now evaluate the expectation value of the three terms in \reff{H1A} in the state $\ket{n,n_z}$:
\beq
\bra{n,n_z}H_x^2\ket{n,n_z}= n^2 \hbar^2 \omega_c^2
\eeq{Hx2} 
\begin{align}
\bra{n,n_z}H_x\,\frac{p_z^2}{m_e}\ket{n,n_z}&= \frac{n \,\hbar \,\omega_c}{m_e}\langle \, p_z^2 \,\rangle\nonumber\\
&= -\frac{1}{2}\hbar^2 \,n\,\omega_c \,\omega_z \langle \, -2a^{\dagger}a-1 \,\rangle\nonumber\\
&=  \hbar^2 \,n\,\omega_c \,\omega_z (n_z + \frac{1}{2})\,.
\end{align}
\begin{align}
\bra{n,n_z}\frac{p_z^4}{4m_e^2}\ket{n,n_z}&=\frac{\hbar^2\omega_z^2}{16}\langle \,a^{\dagger}a^{\dagger}aa+aa^{\dagger}aa^{\dagger}+aaa^{\dagger}a^{\dagger}+a^{\dagger}aa^{\dagger}a+a^{\dagger}aaa^{\dagger}+aa^{\dagger}a^{\dagger}a \,\rangle\nonumber\\
&=\frac{\hbar^2\omega_z^2}{16} (6 n_z^2 +6 n_z+3)
\end{align}
The first order relativistic correction \reff{E1} is then given by 
\beq
E^{(1)}= -\frac{\hbar^2}{2m_e}\Big(n^2 \omega_c^2 + \omega_c \,\omega_z\,n\, (n_z + \frac{1}{2})
 + \frac{\omega_z^2}{16} (6 n_z^2 +6 n_z+3)\Big)\,.
\eeq{E1A}
Note that $E^{(1)}$ is negative and hence lowers the energies. 
\subsection{Energy levels and degeneracy}

\begin{figure}[p]
 \caption{\label{Energy_Levels} The splitting of the energy levels due to the first order relativistic correction. All energies are lowered because the first order correction is negative. For a given $N$, each energy splits into $N+1$ levels for the example considered (where $\omega_c=\omega_z$).}
 \includegraphics[scale=1]{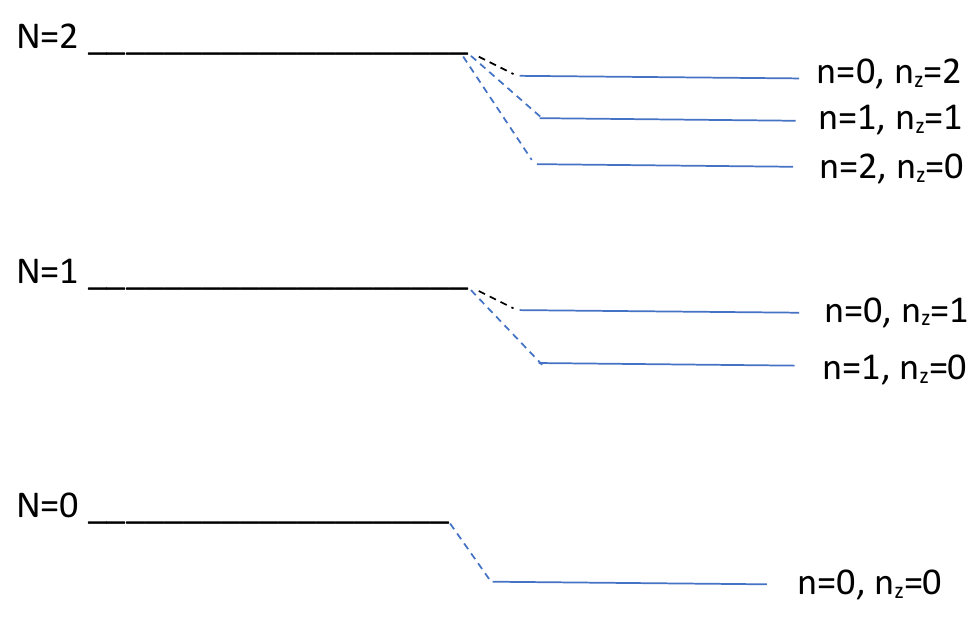}
 \end{figure}

The non-relativistic energy $E^0$ without rest mass term is given by $n \,\hbar \,\omega_c +(n_z+1/2)\, \hbar\omega_z$. As pointed out in \cite{EA}, there is an additional degeneracy besides the usual Landau degeneracy when $\text{w}=\omega_c/\omega_z$ is a rational number. There is an additional degeneracy (as we will later see) when the first order relativistic correction is included but it does not occur any longer at the same $\text{w}$ as in the non-relativistic case. As an illustration that the degeneracy is no longer maintained at the same $\text{w}$ consider the example where $\omega_c=\omega_z$. We then obtain 
\begin{equation}
E^0_{\omega_c=\omega_z}= (n +n_z + 1/2) \hbar\, \omega_c = (N +1/2)\hbar \,\omega_c \quad; N=0,1,2,3,...
\end{equation}
where $N=n+n_z$. There are $N+1$ pairs $(n,n_z)$ with the same value of $N$. Moreover, $N$ pairs out of the $N+1$ have $n \neq 0$ and therefore $N$ pairs have an additional spin degeneracy of $2$. As pointed out in \cite{EA}, the degeneracy due to the $(n,n_z)$ pairs is then $2N+1$. We will refer to this as the $(n,n_z)$ degeneracy to distinguish it from the usual Landau degeneracy $D= \Phi/\Phi_0$ where the magnetic flux $\Phi=B\,A$ and $\Phi_0=h/e$ is a unit of quantum flux. As we will now see, except for the ground state, the first order relativistic correction $E^{(1)}$, given by \reff{E1A},  splits the energy levels so that the $(n,n_z)$ degeneracy at $\omega_c=\omega_z$ is no longer maintained. To illustrate this we will consider the cases $N=0$, $N=1$ and $N=2$. 
\begin{itemize}
\item $N=0$ energy level (the ground state), $E^0= \frac{1}{2}\hbar \,\omega_c$:\\
We have $n=0$ and $n_z=0$. The degeneracy is $2N+1=1$. $E^{(1)}= -\frac{3\,\hbar^2\,\omega_c^2}{32m_e}$. The first order relativistic correction lowers the ground state energy. To get a sense of the magnitude of the correction consider a magnetic field strength of $15T$ which is typical of experiments involving Landau levels \cite{Shubnikov}. Then $E^0= 0.868 \,\text{meV}$ and $E^{(1)}= -0.552 \times 10^{-9} \,\text{meV}$. So the correction is smaller by about a factor of $10^{-9}$. As a comparison, the first order relativistic correction for the hydrogen atom is smaller than the energy by about a factor of $10^{-5}$ \cite{Griffith}.  

\item $N=1$ energy level, $E^0= \frac{3}{2}\hbar \,\omega_c$:\\
We have either $n=0$, $n_z=1$ or $n=1$, $n_z=0$. The degeneracy is $3$.   $E^{(1)}_{n=0,n_z=1}= -\frac{15\,\hbar^2\,\omega_c^2}{32m_e} $ and $E^{(1)}_{n=1,n_z=0}= -\frac{27\,\hbar^2\,\omega_c^2}{32m_e} $. We see that the first order correction splits the $N=1$ energy level into two. $E^{(1)}_{n=0,n_z=1}$ has a spin degeneracy of unity while $E^{(1)}_{n=1,n_z=0}$ has a spin degeneracy of $2$ for a sum of $3$, the degeneracy of the original energy level.

\item $N=2$ energy level, $E^0= \frac{5}{2}\hbar \,\omega_c$:\\
We have either $n=0$, $n_z=2$; $n=1$, $n_z=1$ or $n=2$,$n_z=0$. The degeneracy is $5$.  $E^{(1)}_{n=0,n_z=2}= -\frac{39\,\hbar^2\,\omega_c^2}{32m_e} $, $E^{(1)}_{n=1,n_z=1}= -\frac{55\,\hbar^2\,\omega_c^2}{32m_e}$ and $E^{(1)}_{n=2,n_z=0}= -\frac{83\,\hbar^2\,\omega_c^2}{32m_e} $. We see that the first order correction splits the $N=2$ energy level into three levels. $E^{(1)}_{n=0,n_z=2}$ has a spin degeneracy of unity while $E^{(1)}_{n=1,n_z=1}$ and $E^{(1)}_{n=2,n_z=0}$ each have a spin degeneracy of $2$ for a total sum of $5$, the degeneracy of the original energy level.        
\end{itemize}
The energy level diagram (fig. \ref{Energy_Levels}) depicts the splitting for the first three energy levels. For the above example where $\omega_c=\omega_z$, the first order relativistic correction will split the original energy level into $N+1$ energy levels, $N$ of which will have a spin degeneracy of $2$ and one with a spin degeneracy of unity (for a total sum of $2N+1$).

\begin{figure}[p]
 \caption{\label{Energy_w} Energy E ($E^0 + E^{(1)}$) versus $\text{w}$ where $E$ is in units of $\hbar \omega_z$. There is an $(n,n_z)$ degeneracy at points where lines intersect.   These occur at values of $\text{w}$ that are very slightly shifted from the unperturbed case (see fig. \ref{Zoom1}). The red lines have a spin degeneracy of unity and the blue lines have a spin degeneracy of two.} 
 \includegraphics[scale=1]{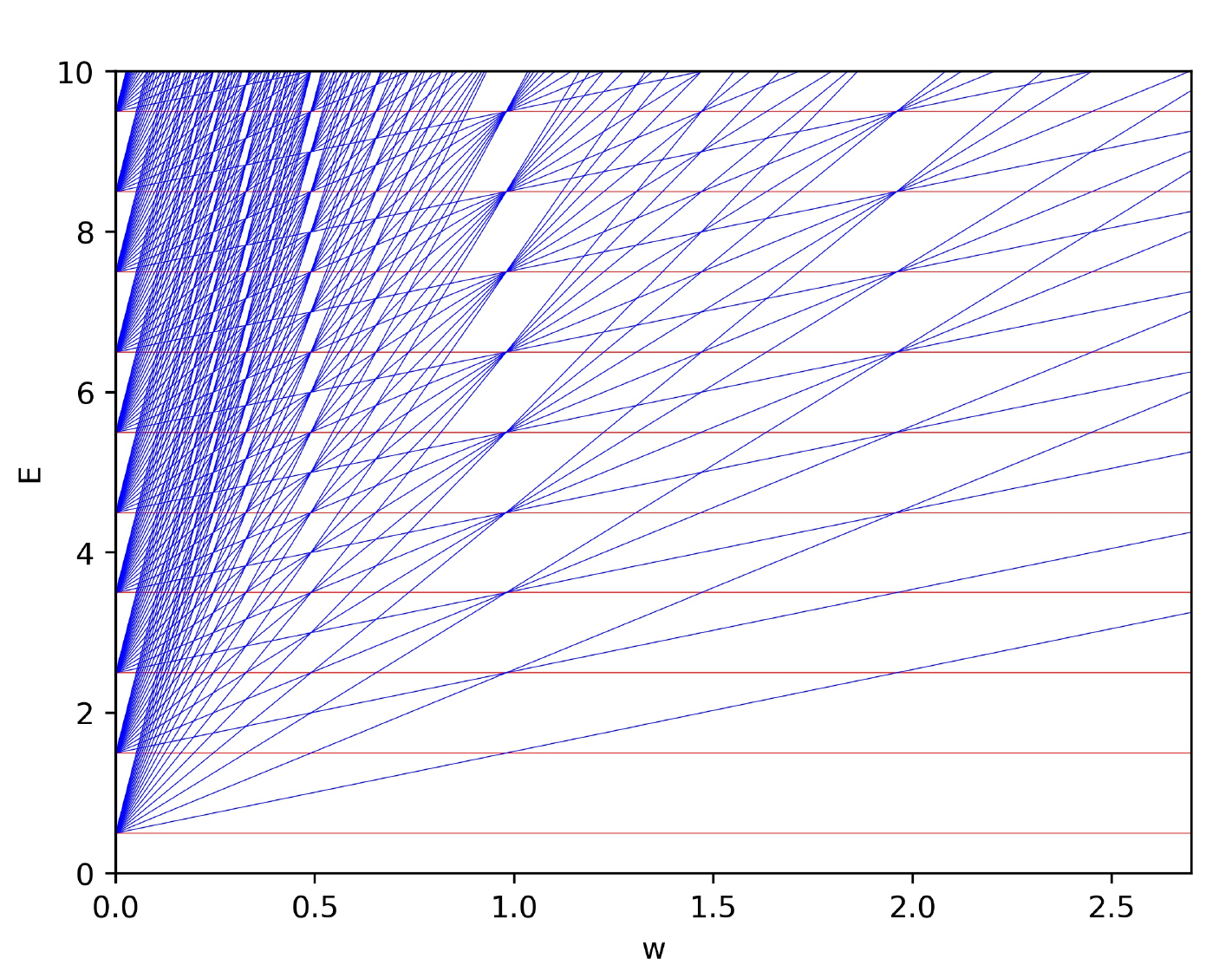}
 \end{figure}

\begin{figure}[p]
 \caption{\label{Zoom1} We zoom in on two regions (shown on the top graph by a circle around two regions). The region around the point of intersection with three lines near $\text{w}=0.98$ is enlarged (graph on bottom right). The region around the point of intersection with five lines near $\text{w}=0.245$ is enlarged (see graph on bottom left). In both cases, we see that the lines do not intersect at one single point in contrast to the non-relativistic case. As before, the red lines have a spin degeneracy of unity and the blue lines have a spin degeneracy of two. For example, at the point of intersection of a red line and a blue line, the degeneracy is $3$ whereas the degeneracy is $4$ at the intersection point of two blue lines. [Note on how to read the bottom graphs: the x-axis on the bottom right hand graph is obtained by multiplying the values on the axis by $10^{-8}$ and then adding $0.98$. For example, the location at $0.5$ is equal to $0.98 + 0.5 \times 10^{-8}$. The y-axis is obtained by multiplying the values on the axis by $10^{-8}$ and then adding 2.5.]}
 \includegraphics[scale=1]{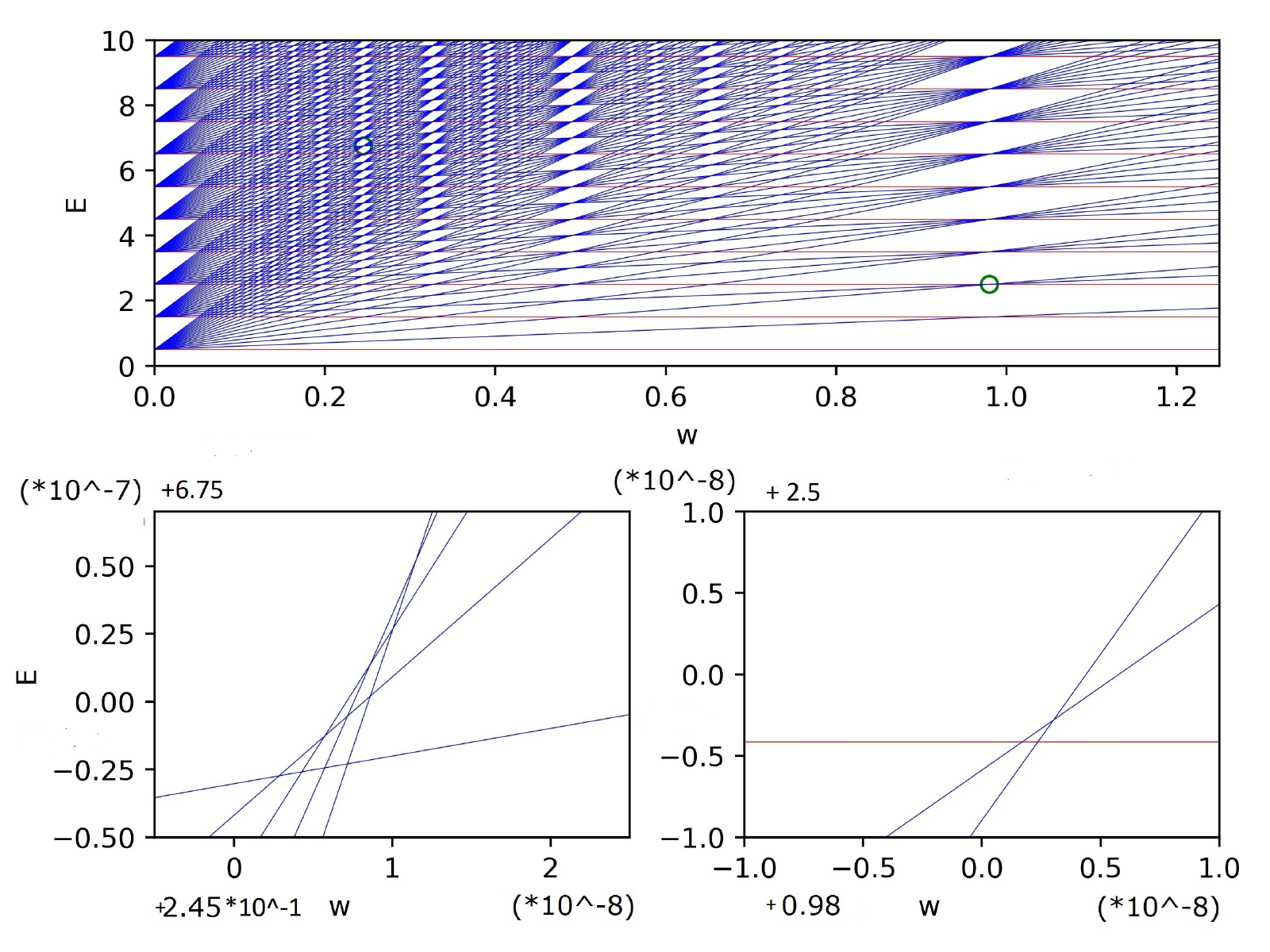}
 \end{figure}

Note that with the first order relativistic correction, an $(n,n_z)$ degeneracy is still present. To see this, we draw a graph of the energy $E=E^0+ E^{(1)}$ vs. $\text{w}$  in fig. \ref{Energy_w} where $E$ is in units of $\hbar \omega_z$. The intersection of lines demonstrates the $(n,n_z)$ degeneracy. To draw this graph we fixed the value of $\omega_z$  to be that of $\omega_c$ when $B=15T$. This yields $\hbar \omega_z/(m_e)$ to be $3.392 \times 10^{-9}$. The smallness of this dimensionless parameter, which captures the smallness of the correction for this choice of $\omega_z$, implies that the graph will be very similar but not identical to the graph of $E^0$ vs. $\text{w}$. That they are not identical can be seen when we zoom in on points of intersection. For example, in fig. \ref{Zoom1} we see on the bottom right hand graph that the three lines near $\text{w}=0.98$ do not meet at one point in contrast to the non-relativistic case where all three lines meet at $\text{w}=1$. Similarly, in fig. \ref{Zoom1}, on the bottom left graph, we see that the five lines near $\text{w}=0.245$ do not meet at one point, in contrast to the non-relativistic case where all five lines meet at $\text{w}=0.25$.  

    
\section{Second order relativistic correction}
The second order relativistic correction has two contributions: one from the expectation value of the second order Hamiltonian $H^{(2)}$ and one from using second order perturbation theory with the first order Hamiltonian $H^{(1)}$ i.e. 
\beq
E^{(2)}= \bra{n,n_z} H^{(2)}\ket{n,n_z} + \sum_{p\neq n_z}\frac{|\bra{n,p}H^{(1)}\ket{n,n_z}|^2}{E^0_{n_z}-E^0_p}\,.
\eeq{E2}
$H^{(2)}$, given by \reff{H2}, can be expressed in terms of $H_x$ and $p_z$,
\begin{align}
H^{(2)}&= \dfrac{1}{2m_e^2}\Big(H_x + \frac{p_z^2}{2m_e}\Big)^3 \nonumber\\
&=\dfrac{1}{2m_e^2}\Big(H_x^3 + 3H_x^2\frac{p_z^2}{2m_e} +3 H_x \frac{p_z^4}{4m_e^2}
+\frac{p_z^6}{8m_e^3}\Big)\,.
\end{align}
We need to evaluate the expectation values of the four terms above in the state $\ket{n,n_z}$, the main parts of which have been calculated in the previous section (except $p_z^6$). The results are
\begin{align}
&\langle \,H_x^3 \,\rangle= n^3 \hbar^3 \omega_c^3\nonumber\\
&\langle \,3H_x^2 \,\frac{p_z^2}{2m_e}\,\rangle=\frac{3}{4}\hbar^3\,\omega_c^2\,\omega_z\,n^2\,(2n_z+1)\nonumber\\
&\langle \, 3 H_x \,\frac{p_z^4}{4m_e^2} \,\rangle=\frac{3}{16}\hbar^3 \,\omega_c\,\omega_z^2\,n \,(6n_z^2+6n_z+3)\nonumber\\
&\langle \,\frac{p_z^6}{8m_e^3} \,\rangle=-\frac{1}{64} \hbar^3\,\omega_z^3 \langle\,(a^{\dagger}-a)^6\,\rangle= -\frac{5}{64} \hbar^3\,\omega_z^3 \,(4n_z^3+6n_z^2+8n_z+3)
\end{align}
where there were twenty terms to evaluate in $\langle\,(a^{\dagger}-a)^6\,\rangle$. 
The first contribution to the second order relativistic correction is therefore given by
\begin{align}
\bra{n,n_z} H^{(2)}\ket{n,n_z}= \dfrac{\hbar^3}{2m_e^2}\Big(n^3 \omega_c^3 & +\frac{3}{4}\omega_c^2\,\omega_z\,n^2\,(2n_z+1) \nonumber\\ & +\frac{3}{16}\,\omega_c\,\omega_z^2\,n\,(6n_z^2+6n_z+3)
-\frac{5}{64}\,\omega_z^3 \,(4n_z^3+6n_z^2+8n_z+3) \Big)\,.
\label{H2B}
\end{align}
We now evaluate the second contribution to $E^{(2)}$ using $H^{(1)}$ given by \reff{H1A}. There are four separate cases to evaluate.

Case 1: $p=n_z-4$\\
\begin{align}
\frac{|\bra{n,n_z-4}H^{(1)}\ket{n,n_z}|^2}{E^0_{n_z}-E^0_{n_z-4}}&= \Big(\frac{\hbar^2\omega_z^2}{32 m_e}\Big)^2 \frac{|\bra{n,n_z-4}a\,a\,a\,a\ket{n,n_z}|^2}{4 \hbar \omega_z}
\nonumber\\&= \frac{\hbar^3 \omega_z^3}{4096\,m_e^2} n_z(n_z-1)(n_z-2)(n_z-3)
\end{align}
Case 2: $p=n_z+4$\\
\begin{align}
\frac{|\bra{n,n_z+4}H^{(1)}\ket{n,n_z}|^2}{E^0_{n_z}-E^0_{n_z+4}}&= \Big(\frac{\hbar^2\omega_z^2}{32 m_e}\Big)^2 \frac{|\bra{n,n_z+4}a^{\dagger}\,a^{\dagger}\,a^{\dagger}\,a^{\dagger}\ket{n,n_z}|^2}{(-4 \hbar \omega_z)}
\nonumber\\&= -\frac{\hbar^3 \omega_z^3}{4096\,m_e^2}\, (n_z+1)(n_z+2)(n_z+3)(n_z+4)
\end{align}
Case 3: $p=n_z+2 $\\
\begin{align}
\frac{|\bra{n,n_z+2}H^{(1)}\ket{n,n_z}|^2}{E^0_{n_z}-E^0_{n_z+2}}&= \frac{1} {(-2 \hbar \omega_z)}\left |\frac{\hbar\omega_z}{4 m_e} \bra{n,n_z+2}H_x\, a^{\dagger}\,a^{\dagger}\ket{n,n_z}\right. \nonumber\\&\left.+\frac{\hbar^2 \omega_z^2}{32m_e}\bra{n,n_z+2} a^{\dagger}\,a^{\dagger}\,a^{\dagger}\,a+a^{\dagger}\,a^{\dagger}\,a\,a^{\dagger}+a^{\dagger}\,a\,a^{\dagger}\,a^{\dagger}+a\,a^{\dagger}\,a^{\dagger}\,a^{\dagger}\ket{n,n_z}\right|^2 \nonumber\\&=-\frac{\hbar^3}{512 \,m_e^2} \omega_z(n_z+1)(n_z+2) (4 \,
\omega_c \,n +2\, n_z\,\omega_z+3 \,\omega_z)^2 
\end{align}
Case 4: $p=n_z-2$\\
\begin{align}
\frac{|\bra{n,n_z-2}H^{(1)}\ket{n,n_z}|^2}{E^0_{n_z}-E^0_{n_z-2}}&= \frac{1} {2 \hbar \omega_z}\left|\frac{\hbar\omega_z}{4 m_e} \bra{n,n_z-2}H_x\, a\,a\ket{n,n_z} \right.\nonumber\\&\left.+\frac{\hbar^2 \omega_z^2}{32m_e}\bra{n,n_z-2} a^{\dagger}\,a\,a\,a+a\,a^{\dagger}\,a\,a + a\,a\,a^{\dagger}\,a + a\,a\,a\,a^{\dagger}\ket{n,n_z}\right|^2 \nonumber\\&=\frac{\hbar^3}{512 \,m_e^2} \omega_z\,n_z(n_z-1) (4 \,\omega_c \,n +2\, n_z\,\omega_z-\omega_z)^2 
\end{align}
The second contribution to $E^{(2)}$ is the sum of the four above cases:
\begin{align}
\sum_{p\neq n_z}\frac{|\bra{n,p}H^{(1)}\ket{n,n_z}|^2}{E^0_{n_z}-E^0_p}&= -\dfrac{\hbar^3 \omega_z}{512 \,m_e^2}\Big(32\,\omega_c^2 \,n^2(2n_z+1)\nonumber\\& +48\,\omega_c\,\omega_z\,n(2\,n_z^2+2\,n_z+1)+ \omega_z^2\,(34\, n_z^3 +51 \,n_z^2 +59\,n_z +21)\Big)\,.
\label{E22}
\end{align}
The second order relativistic correction is then the sum of \reff{E22} and \reff{H2B}:
\begin{align}
E^{(2)}=\dfrac{\hbar^3}{2\,m_e^2}\Big[n^3\,\omega_c^3 &+\dfrac{5}{8} n^2 \,\omega_c^2\,\omega_z (2 n_z+1) \nonumber\\&+\dfrac{1}{8} n\,\omega_c\,\omega_z^2 (6n_z^2 +6 n_z +3)\nonumber\\&-\dfrac{1}{256} \,\omega_z^3 (114 n_z^3 +171 n_z^2 +219 n_z +81)\Big]\,.
\label{E2F}
\end{align}
\section{Conclusion}
The main results of this paper are the compact formulas \reff{E1A} and \reff{E2F} for the first and second order relativistic corrections to the energy respectively. These are relativistic corrections for an electron moving under a constant magnetic field and a parallel linear electric field.  In our previous non-relativistic work \cite{EA} we had shown that besides the Landau degeneracy, the presence of the parallel linear electric field adds an extra $(n,n_z)$ degeneracy when the ratio $\text{w}=\omega_c/\omega_z$ is a rational number. With the first order relativistic correction, we again obtain an $(n,n_z)$ degeneracy but it no longer occurs at the same value of $\text{w}$ as in the non-relativistic case (it is slightly shifted from the unperturbed value). That it no longer occurs at the same $\text{w}$ was illustrated in the energy level diagram fig.\ref{Energy_Levels} where we observe the splitting of the energy levels for the case $\omega_c=\omega_z$. With the inclusion of the first order relativistic correction, fig.\ref{Energy_w} shows that there is clearly an $(n,n_z)$ degeneracy at points where lines intersect. Figure \ref{Energy_w} is very similar to the graph in the non-relativistic case but not identical. This can be seen from the fact that when we zoom in on points of intersection (see the bottom graphs of fig. \ref{Zoom1}) the lines do not actually all meet at one point (whereas they do in the non-relativistic case). 

A ring of charge placed in the x-y plane generates a linear electric field in the 
$z$-direction in the vicinity of its center. However, it also produces an electric field in the x-y plane (see \cite{EA} for details). For future work, it would be worthwhile and interesting to calculate the relativistic corrections for this experimental ring set-up where an extra electric field in the x-y plane is present. It is expected in this case that the Landau degeneracy itself will be lifted and be replaced by tightly spaced ``bands".

\section*{Acknowledgments}
A.E. acknowledges support from a discovery grant of the National Science and Engineering Research Council of Canada (NSERC).

\end{document}